\documentclass[letter]{aa} 
\newcommand{\kms}{\mbox{km~s$^{-1}$}}
\newcommand{\ic}{IC\,1396\,N}
\newcommand{\rf}{}
\usepackage{graphicx}
\usepackage{txfonts}
\begin{document}
   \title{The IC1396N proto-cluster at a scale of $\sim$250~AU
\thanks{\,Based on observations obtained at the IRAM Plateau de Bure
Interferometer (PdBI). IRAM is funded by the Centre Nationale de la
Recherche Scientifique (France), the Max-Planck Gesellschaft
(Germany), and the Instituto Geogr\'afico Nacional (Spain).}}
   \author{R. Neri\inst{1}
          \and
          A. Fuente\inst{2}
          \and
          C. Ceccarelli\inst{3}
          \and
          P. Caselli\inst{4,5}
          \and
          D. Johnstone\inst{6,7}
          \and
          E.F. van Dishoeck\inst{8}
          \and
         F. Wyrowski\inst{9}			
          \and
          M. Tafalla \inst{2}
          \and
          B. Lefloch \inst{3}
          \and
          R. Plume \inst{9}
          }
 
   \offprints{R. Neri -- e-mail: neri@iram.fr}

   \institute{Institut de Radioastronomie Millim\'etrique, 300 rue de la Piscine, 38406 St Martin d'H\`eres Cedex, France
             \and
             Observatorio Astron\'omico Nacional (OAN), Apdo. 112,
            E-28803 Alcal\'a de Henares (Madrid), Spain 
             \and
             Laboratoire d'Astrophysique de l'Observatoire de Grenoble, BP 53, 38041 Grenoble Cedex 9, France
              \and
              Osservatorio Astrofisico di Arcetri (INAF), Largo E.\ Fermi 5, 50125 Firenze, Italy 
              \and
              Harvard-Smithsonian Center for Astrophysics, 60 Garden Street, Cambridge, MA 0213
              \and
               Department of Physics and Astronomy, University of Victoria, Victoria, BC V8P 1A1, Canada
               \and		
               National Research Council of Canada, Herzberg Institute, 5071 West Saanich Road, Victoria, BC V9E 2E7, Canada 
              \and
              Leiden Observatory, PO Box 9513, 2300 RA Leiden, Netherlands
             \and
             Max-Planck-Institut f\"ur Radioastronomie, Auf dem H\"ugel 69, 53121 Bonn, Germany 
             \and
             University of Calgary, 2500 University Drive NW, Alberta  T2N 1N4, Canada}
   \date{Received February 17, 2007; accepted April 12, 2007}

 
\abstract
{} 
{\rf We investigate the mm-morphology of \ic\
with unprecedented spatial resolution to analyze its dust and
molecular gas properties, and draw comparisons with objects of similar
mass.}
{\rf We have carried out sensitive observations in the most extended
configurations of the IRAM Plateau de Bure interferometer, to map the
thermal dust emission at 3.3 and 1.3mm, and the emission from the
$J$=13$_k\rightarrow$12$_k$ hyperfine transitions of methyl cyanide
(CH$_3$CN).}
{\rf We unveil the existence of a sub-cluster of hot cores in
\ic, distributed in a direction perpendicular to the emanating
outflow. The cores are embedded in a common envelope of extended and
diffuse dust emission. We find striking differences in the dust
properties of the cores ($\beta\simeq$\,0) and the surrounding
envelope ($\beta\simeq$\,1), very likely testifying to differences in
the formation and processing of dust material. The CH$_3$CN emission
peaks towards the most massive hot core and is marginally extended in
the outflow direction.}
{}


   \maketitle
%

\section{Introduction}

Intermediate-mass young stellar objects (IMs) (protostars and Herbig
Ae/Be stars with M$_\star$ $\sim$2$-$10\,M$_{\odot}$) are crucial to
star formation studies because they provide the link between
evolutionary scenarios of low- and high-mass stars.  These objects
share many similarities with high-mass stars (clustering, PDRs).
However, to study them presents decided advantages, compared to
massive star forming regions, as many of them are located close to the
Sun ($\leq$1\,kpc) and in regions of reduced complexity.

The lack of spatial resolution is always a potential source of
confusion in the identification of isolated stellar structures. This
issue becomes increasingly important in the case of low-mass and
intermediate-mass stars, which are predisposed to form in
clusters. While much is now known about isolated star formation, the
onset of cluster formation remains less clear and systematic surveys
are needed to gain a deeper understanding (Testi et al.\
2000). Aperture synthesis in the millimeter range provides excellent
means to investigate and characterize the earliest stellar clustering.

In this Letter, we present interferometric continuum and molecular
line observations of the IM protostar \ic\ at the highest spatial
resolution of the PdBI (Karastergiou et al.\ 2006). \ic\ is a
$\sim$440\,L$_\odot$ (Sugitani et al.\ 2000) protostar located at a
distance of 750\,pc (Matthews 1979). Classified as a Class 0\,/\,I
borderline source, this young protostar is associated with a very
energetic bipolar outflow (Codella et al.\ 2001). In addition,
near-infrared images by Nisini et al.\ (2001) have revealed the
presence of a collimated jet.  Thus far, it is one of the best studied
Class 0\,/\,I intermediate-mass sources.


\section{Observations}
We used the six-element IRAM Plateau de Bure Interferometer to observe
\ic\ simultaneously in three hyperfine transitions of CH$_3$CN
($J$=13$\rightarrow$12) at 239.064\,GHz, in the A- and E-state
transitions of CH$_3$OCHO ($J$=7$\rightarrow$6) and
($J$=8$\rightarrow$7) at 90.188\,GHz, and in the continuum at 1.3mm
and 3.3mm. Observations were made in configuration A (January 27,
2005), new A (February 12, 2006) and in a non-standard configuration
(March 22, 2006) in conditions of excellent atmospheric seeing
(0.3$-$0.5$''$ FWHM) and good atmospheric transparency (pwv =
1$-$4\,mm). The spectral correlator was adjusted to observe the line
transitions with a velocity resolution of 0.3\,\kms\ and to cover the
entire RF passbands (580\,MHz) for highest continuum sensitivity. The
overall flux density scales for each epoch and for each frequency band
were set by comparison to the known flux densities of MWC349 and 3C273
and estimated to be accurate to 10\% at 3.3mm and 20\% at 1.3mm. The
resulting line and continuum point source sensitivities were estimated
to 8.4\,mJy (2.5\,MHz) and 0.43\,mJy at 237.571\,GHz, and to 2.5\,mJy
(2.5\,MHz) and 0.12\,mJy at 90.250\,GHz. The corresponding synthesized
beams adopting uniform weighting were 0.45$''\times0.37''$ at
PA=$-9$\degr\ and 1.2$''\times1.0''$ at PA=$-11$\degr. The data were
calibrated, mapped and analyzed in the GILDAS software package.
\begin{table*}
\caption{Millimeter flux densities, sizes, spectral indices and masses}             
\label{table:1}      
{\centering          
\begin{tabular}{l c c c c c }     
\hline\hline       
\\
\multicolumn{1}{l}{\rf 3.3mm (90.3 GHz)} 
& \multicolumn{1} {c}{BIMA 3}  &  \multicolumn{4}{c}{-------------------------- BIMA 2 --------------------------} \\
                  &             &  Cocoon      & 41.86+11.9   &  41.73+12.8   &  41.73+14.3  \\ \hline
$\alpha$ (J2000)  & 21$^{\rm h}$40$^{\rm m}$42\fs84 & 21$^{\rm h}$40$^{\rm m}$41\fs86  & 21$^{\rm h}$40$^{\rm m}$41\fs85  &  21$^{\rm h}$40$^{\rm m}$41\fs73  & 21$^{\rm h}$40$^{\rm m}$41\fs72  \\
$\delta$ (J2000)  & 58\degr16\arcmin01\farcs4  & 58\degr16\arcmin13\farcs2   & 58\degr16\arcmin11\farcs9   &  58\degr16\arcmin12\farcs8   &  58\degr16\arcmin14\farcs3  \\
Size ($"$)        &  $0.8'' \times 0.5''$  &  $4.3'' \times 3.1''$   & $\le 0.5''$   & $\le 0.5''$    & $\le 0.3''$  \\        
S (mJy)           &      8       &     16       &     5.9      &      1.5      &         2.7  \\ \hline 
\\
\multicolumn{1}{l}{\rf 1.3mm (237.6 GHz)} 
& \multicolumn{1} {c}{BIMA 3}  &  \multicolumn{4}{c}{-------------------------- BIMA 2 --------------------------}\\
                                 &     &  Cocoon    &    41.86+11.9   &  41.73+12.8    &  41.73+14.3  \\ \hline
$\alpha$ (J2000)  &   21$^{\rm h}$40$^{\rm m}$42\fs84      &   21$^{\rm h}$40$^{\rm m}$41\fs85      & 21$^{\rm h}$40$^{\rm m}$41\fs86        &   21$^{\rm h}$40$^{\rm m}$41\fs73  &  21$^{\rm h}$40$^{\rm m}$41\fs73  \\ 
$\delta$ (J2000)  &   58\degr16\arcmin01\farcs4       &   58\degr16\arcmin13\farcs1       & 58\degr16\arcmin11\farcs9         &   58\degr16\arcmin12\farcs8   &  58\degr16\arcmin14\farcs3   \\
Size ($"$)        & $0.8''\times0.4''$ & $4.5''\times3.1''$ & $0.4''\times0.2''$ & $\le 0.3''$    &   $\le 0.2''$ \\
S (mJy)           &        30          &       245          &       35           &          6     &     10        \\ \hline
\\
\multicolumn{1}{c}{} & \multicolumn{1} {c}{BIMA 3}  &  \multicolumn{4}{c}{-------------------------- BIMA 2 --------------------------}\\
                                 &     &  Cocoon    &    41.86+11.9   &  41.73+12.8    &  41.73+14.3  \\ \hline
Mean Spec.Index   &      1.4     &    \multicolumn{4}{c}{1.9} \\
Spec.Index ($\alpha$)  &      1.4     &         2.8          &       1.9       &         1.5        &      1.4 \\
Mass (M$_\odot$)$^a$  &     0.05     &         0.4          &       0.06      &       0.01         &     0.01   \\
$\beta=\alpha-2$  &     $\le0.0$ &          0.8     &      $-0.1$   &      $\le0.0$      &  $\le0.0$  \\
\hline                  
\end{tabular}}\\

{\noindent \rf $^a$ masses have to be scaled up by a factor of $\sim$6 for
a dust temperature of 20\,K.}
\end{table*}

\section{Continuum data}
Figures 1a and 1b show the continuum images obtained from the
visibility data in the two frequency bands. The high sensitivity and
the striking similarity of the many morphological features in the two
figures, clearly testify to the presence of at least three bright
continuum emission cores in the center of \ic, all three very likely
associated with the centrally peaked dust emission source identified
as BIMA 2 by Beltr\'an et al.\ (2002). While the two weaker cores were
not resolved by the interferometer, the primary core IRAM{\rf\,2A} is
resolved in the 1.3mm emission to an elliptical region of
$\sim$300\,AU$\times 150$\,AU. This corresponds to $5\times$ the size
of the solar-system, and may be an accretion disk surrounding the
powering source of the bipolar outflows mapped by Beltr\'an et al.\
(2004) in various molecular transitions. The bulk of the mm-emission,
however, is emerging from a much larger region centered on the triple
system and oriented almost perpendicular to the molecular outflows
(see Beltr\'an et al.\ 2002).  This {\rf oval-shaped} region whose
extent ($\sim$2800\,AU) is very much the same at both observing
frequencies, provides about 50\% of the 1.3mm flux density detected at
the JCMT (Saraceno et al.\ 1996).

According to these results, we consider two different models for the
continuum emission: a large volume of cold dust (resolved out by the
PdBI) that surrounds (a) an envelope with sharp boundaries in which
the three compact cores are embedded, (b) a confined region harboring
multiple lower brightness cores from which we have detected the three
most intense.  Though the lack of sensitivity to large scale emission
and emission distributed over a large number of cores makes it
difficult to argue against one or the other model, we favor the
former, more simple model of the dusty `cocoon' {\rf at the border of
which} the three intense cores are embedded. We obtain excellent
results by fitting such a model at 3.3mm and 1.3mm, which strengthens
the `cocoon+cores' interpretation for the protostellar region {\rf
associated} with BIMA 2. The fitted parameters of our model and the
corresponding model image are shown in Table 1 and Figure 1 (bottom).
We cannot, however, totally discard the possibility of the existence
of a relatively homogeneous cluster of lower mass cores surrounding
the three bright cores detected.

In addition to the intense clump BIMA 2, Beltr\'an et al.\ (2002)
detected two clumps in continuum emission, BIMA 1 and BIMA 3.  In
spite of the much higher sensitivity of the observations presented
here, BIMA 1, the clump in the south-west of BIMA 2, was not detected
in the 3.3mm continuum image. It is likely that the interferometer
resolved out extended ($\le 5''$) low brightness continuum emission
from this source. BIMA 3, the radio-bright source in the south-east of
BIMA 2, is detected at the edge of our continuum images. The continuum
emission towards BIMA 3 can be interpreted as compact emission
emerging from a region $\sim$450\,AU in size, embedded in a larger
cloud of warm and dense dust and resolved out by the interferometer,
accounting for about 70\% of the continuum emission losses at 1.3mm.

   \begin{figure}
   \centering
    \includegraphics[angle=0,width=8.2cm]{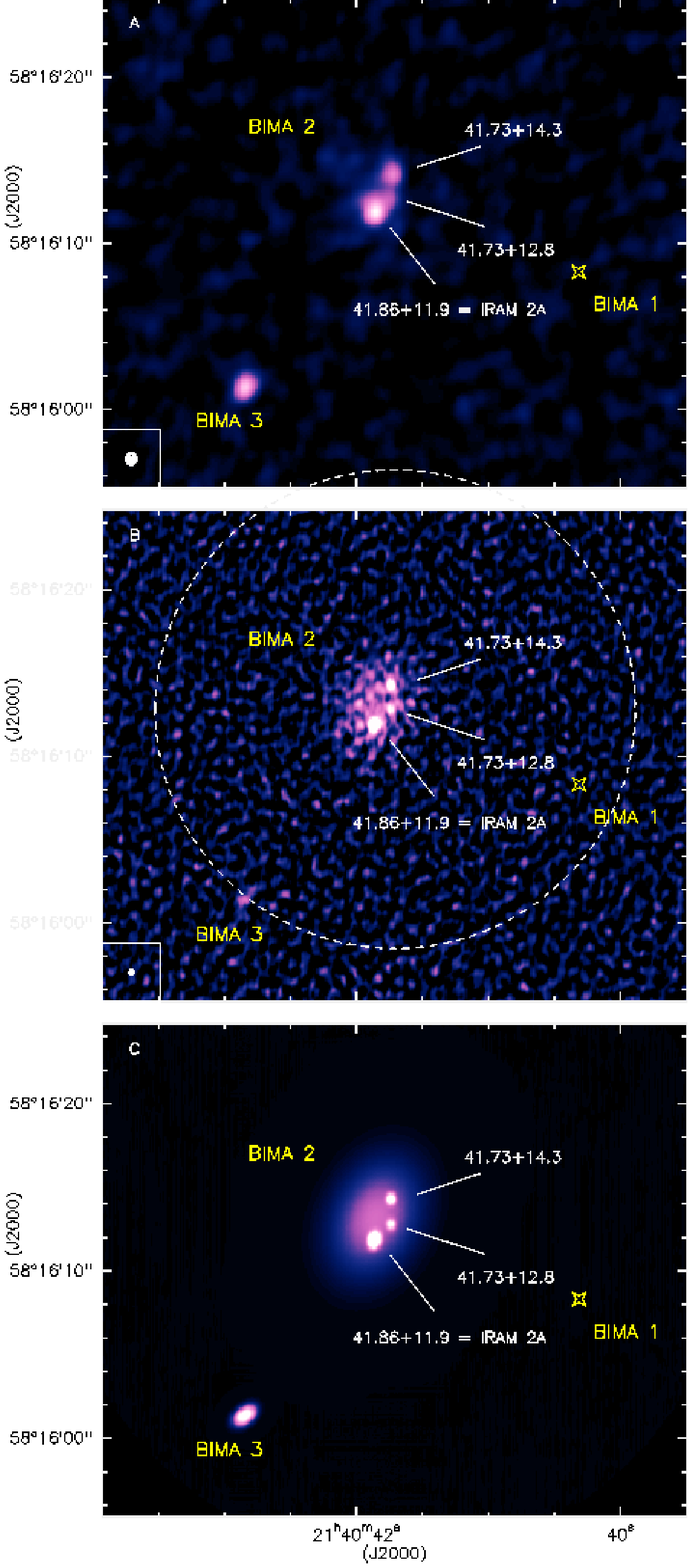}
      \caption{Continuum map of \ic\ at 90.250\,GHz ({\em top}) and
	237.570\,GHz ({\em middle}). The $1\sigma$ noise levels are
	0.12{\rf\,mJy/beam (0.02 K)} and 0.43{\rf\,mJy/beam (0.06 K)},
	respectively for the 3.3mm and 1.3mm continuum. The ellipse in
	the lower left corners shows the synthesized beam, the dashed
	ellipse in the middle panels shows the FOV of the 1.3mm
	receivers. The center of the dashed ellipse marks the position
	of the phase center at $\alpha = 21^{\rm h}$40$^{\rm
	  m}$41\fs71 and $\delta=58\degr16\arcmin12\farcs8$
	(J2000). The figure ({\em bottom}) shows the 1.3mm
	reconstructed and primary beam corrected model image. BIMA 1
	({\em open star}) is to the SW of BIMA 2. }
         \label{Fig 2}
  \end{figure}

   \begin{figure}
   \centering
   \includegraphics[angle=-90,width=9cm]{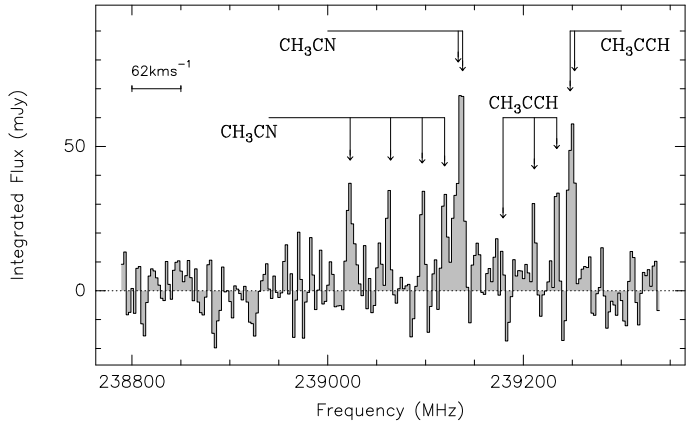}
      \caption{CH$_3$CN (13$_k\rightarrow12_k, k=0\ldots5$) and
      CH$_3$CCH (14$_k\rightarrow13_k, k=0\ldots4$) lines at 239.064
      GHz observed with the PdBI towards the core 41.86+11.9 (=
      IRAM{\rf\,2A}). The intensity scale is in units of mJy and the
      spectral resolution is 2.5\,MHz (= 3.1\,km\,s$^{-1}$). The one
      sigma noise is 8.4\,mJy.}
         \label{Fig 3}
   \end{figure}

   \begin{figure}
   \centering
   \includegraphics[angle=-90,width=8cm]{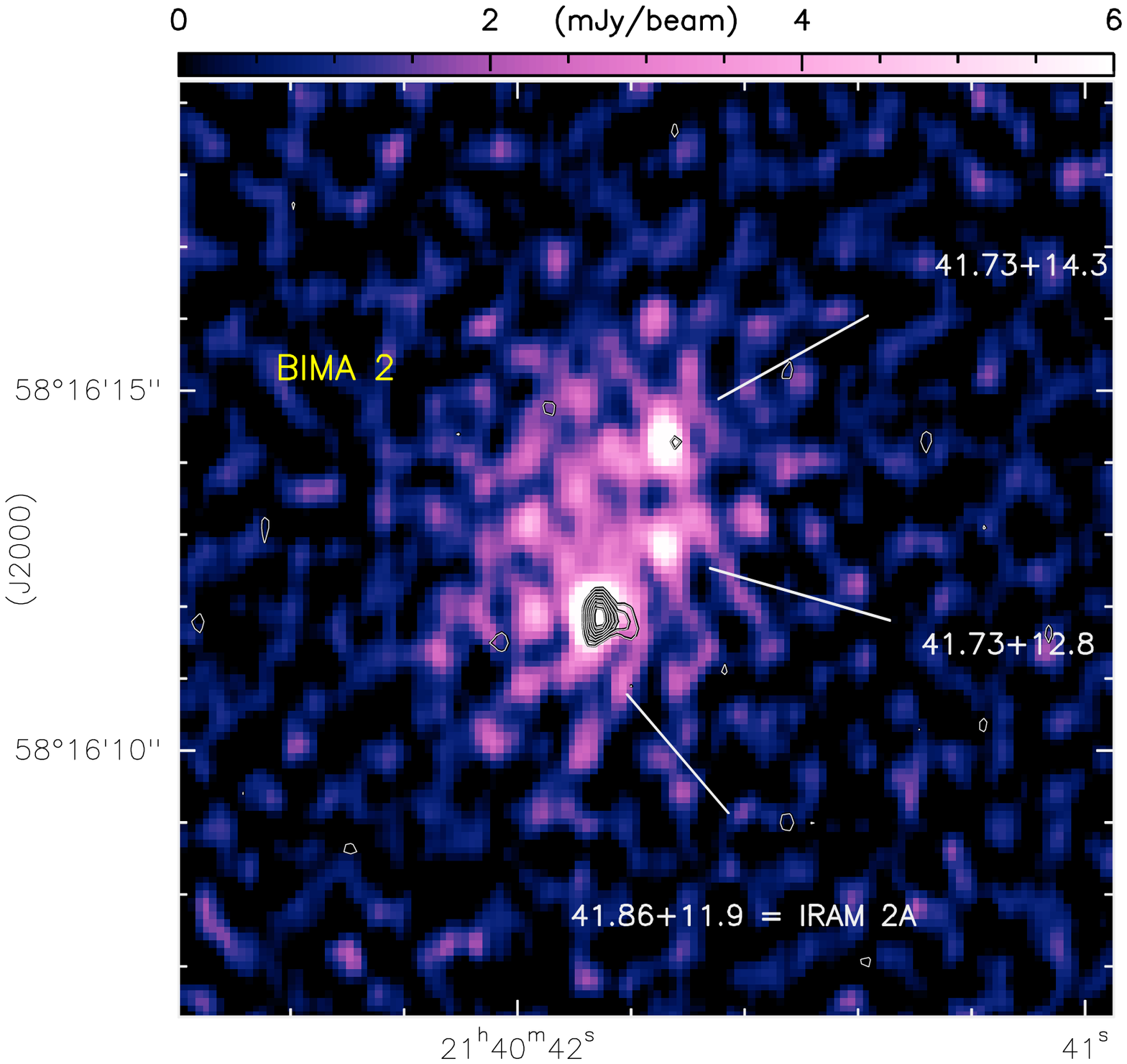}
      \caption{Continuum emission at 1.3mm superposed on the map of
the integrated intensity flux of the $k$\,=\,0,1,2 components of the
CH$_3$CN 13$_k$$\rightarrow$12$_k$ line (contours). Levels for the
continuum emission are from 3$\sigma$ to 9$\sigma$ in steps of
1$\sigma$\,=\,0.12\,Jy~km~s$^{-1}$. Significant CH$_3$CN emission is
detected towards the most massive core only.}
         \label{Fig 1}
   \end{figure}

\subsection{Spectral Indices and Masses}

In Table 1 we present the primary beam corrected flux densities, {\rf
sizes} and spectral indices derived by fitting our model to the
visibilities in the uv-plane. The spectral indices range from {\rf
2.8} for the cocoon to {\rf $\sim$1.4--1.9} for the compact
cores. While the spectral index for the cocoon is consistent with
optically thin dust emission with $\beta\sim1$, the value measured
towards any of the compact cores is very low ($\alpha<2$).

Different explanations may account for the low spectral
index measured towards the compact cores. One possibility is that the
flux densities are a mix of dust and free-free emission.  Beltr\'an et
al.\ (2002) reported VLA continuum flux densities at 3.6cm and upper
limits at 12cm for BIMA 2 and BIMA 3. Since BIMA 2 presents a complex
structure at the high angular resolution of our observations, we
compare the flux densities at centimeter and millimeter wavelengths
in BIMA 3 only.  We have extrapolated the emission at 3.6cm to
millimeter wavelengths assuming the spectral index of an isotropic
ionized wind ($S_\nu\propto\nu^{0.6}$) and subtracted it from the PdBI
flux densities. The resulting values were then fitted assuming
optically thin dust emission. The ratio between the
1.3mm and 3.3mm emission can then only be fitted if we assume a spectral
index of $\beta\sim0$ for the dust opacity.  One possibility is that
the dust emission is optically thick, which is, however, not 
consistent with the measured flux densities. With $\kappa_{\rm 1.3mm}
= 0.01$cm$^{-2}$g$^{-1}$, a gas and dust mass of $\sim${\rf
1.1}~M$_\odot$ is needed to yield opacities $>$1 in BIMA 3.  A
moderately warm core (T$_{\rm dust}$$\sim$50~K) of {\rf
1.1}~M$_{\odot}$ would emit a flux density of {\rf 300}~mJy at 1.3mm,
i.e.\ a factor of 10 more than observed. Although our observations are
not consistent with an optically thick core of $\sim$450~AU, {\rf and
despite a relatively low surface brightness temperature ($\sim$4\,K)
at 1.3mm}, we cannot discard the possibility that part of the core is
optically thick. Clearly, an optically thick region would contribute
to a lower value of the spectral index. 

Another possibility is to assume a different distribution of grain
type in the compact cores and in the surrounding cocoon. Low values of
$\beta$ ($\sim$0-0.5) are found in circumstellar disks around Herbig
Ae and Be stars (Natta et al.\ {\rf 2004, 2007}; Fuente et al.\ 2003,
2006) and have been interpreted as evidence of grain growth. According
to this interpretation, the differences in the spectral index we
measure in the cocoon and in the compact sources could hint at
differences in the grain properties at different spatial scales.
These call for a scenario in which the compact components are
circumstellar disks embedded in a dust-enshrouded enviroment. {\rf
Similar differences in the spectral index between the envelope and the
compact components of Class 0 sources were reported by J{\o}rgensen et
al.\ (2007).}

To derive mass estimates for the different components (see Table 1),
we have assumed a dust temperature of 100\,K for the cocoon and the
cores. Although the dust grain properties are very likely to be
different, we have based our mass estimates on a uniform dust
emissivity index $\kappa_{1.3mm} = 0.01$cm$^{-2}$g$^{-1}$. All the
compact cores have masses much lower than one solar mass, possibly
suggesting that the PdBI has resolved out a large fraction of the
envelope. We can compare the mass of the cores with those derived
towards hot corinos and intermediate mass hot cores.  After correcting
for the distance, the 1.3mm flux density towards IRAM{\rf\,2A} is a
factor of 3$-$4 larger than the one measured towards the hot corinos
IRAS 16293--2422 A/B (Bottinelli et al.\ 2004a) and testifies to the
presence of a more massive star. However, IRAM{\rf\,2A} is 10$\times$
weaker than the intermediate mass hot core NGC~7129--FIRS~2 suggesting
that IRAM{\rf\,2A} is less massive, although both protostars are
located at {\rf about} the same distance. Despite the fact that \ic\
has a luminosity similar to NGC~7129--FIRS~2, the envelopes
surrounding these stars very likely correspond to very different
star formation regimes. While NGC~7129--FIRS~2 at prima facie hosts a
single massive core (pending higher resolution observations), \ic\
definitely hosts a cluster of lower mass stars. We argue that the most
massive component of the \ic\ cluster is very likely the precursor of
a Herbig Ae star.

%

\section{Molecular line data: CH$_3$CN}

CH$_3$CN (13$_k$$\rightarrow$12$_k$) line emission has only been
detected towards IRAM{\rf\,2A}, the most massive core in the \ic\
proto-cluster. In addition to the CH$_3$CN lines, several CH$_3$C$_2$H
lines appear in the spectrum testifying to the presence of a core,
rich in complex molecules (see Fig.\ 2). The integrated intensity map
of the CH$_3$CN (13$_k$$\rightarrow$12$_k$) transition shows that the
emission arises in a region of radius $\sim$0.8$''$ (600\,AU) around
the massive hot core IRAM{\rf\,2A} and with a good positional
coincidence between line and 1.3mm continuum emission. 

Comparing the integrated emission with the peak flux density, we
estimate that $\sim$50\% of the emission is coming from a compact
region (radius $\le$150\,AU) around the hot core while the remaining
50\% arises from a more extended and cooler component in the immediate
vicinity. Bottinelli et al.\ (2007) report about a similar case in
NGC1333\,IRAS\,4A. Although the attained sensitivity is not deep
enough to draw firm conclusions on the small-scale spatial
distribution of the CH$_3$CN emission, the shape of the emission
suggests that CH$_3$CN is tracing the signature of gas and dust {\rf
in the hot core and possibly arises from a narrow boundary layer at the
outflow-core interface}. The marginal or non-detection of CH$_3$CN
towards the weaker cores is possibly subject to the same sensitivity
limitation.

We have tried to fit the rotational diagram of the CH$_3$CN
(13$_k$$\rightarrow$12$_k$) transition towards the emission peak. The
K-ladder cannot be fitted with a single rotational temperature
suggesting that the CH$_3$CN emission is optically thick in the hot
core. Also, we cannot exclude that some K-components are blended with
other lines. Given all the uncertainties, we derive a CH$_3$CN
abundance using the temperature inferred from the rotational diagram.
We obtain a rotational temperature of 240$\pm$150~K and
$N$(CH$_3$CN)$=7.1\,10^{14}$ cm$^{-2}$ towards the peak, which implies
a CH$_3$CN abundance of 2\,10$^{-10}$. This abundance is similar to
the ones determined by Bottinelli et al.\ (2004b) in the hot corinos
NGC1333\,IRAS\,4A and IRAS\,16293--2422 but a factor of 10 lower than
the one measured by Fuente et al.\ (2005) towards
NGC~7129--FIRS~2. This low abundance of CH$_3$CN also argues in favor
of \ic\ being a cluster of low\,/\,intermediate mass stars.

\section{Summary}

We report on interferometric 3mm and 1.3mm continuum and CH$_3$CN
(13$_k$$\rightarrow$12$_k$) line observations of the intermediate-mass
protostar \ic\ in the most extended configuration of the PdBI. Our
observations provide a spatial resolution of $\sim$250\,AU at the
distance of \ic\ and reveal the existence of a cluster of cores in
this IM protostar with unprecedented detail and quality. The cluster
in the interior of \ic\ comprises at least three cores, the very
likely precursors of two low-mass and an IM star. Because of
sensitivity limitations, we cannot exclude the presence of further low
brightness cores in the BIMA\,2 region. Our continuum observations
also show that the spectral index at millimeter wavelenghts changes
with the spatial scales traced by the interferometer. The variations
in the spectral index are very likely driven by grain properties
changing with gas density and by the existence of compact regions in
the hot cores with optically thick emission.

CH$_3$CN (13$\rightarrow$12) line emission has only been detected towards
the most massive core IRAM{\rf\,2A}, but our sensitivity limit is not
low enough to infer a different abundance in the other cores.  The
CH$_3$CN emission is slightly more extended than the associated dust
emission and possibly arises from a region heated up by the
molecular outflow.

\begin{acknowledgement}
{\rf We would like to thank the referee for her/his constructive
remarks}. A.F.\ is grateful for support from the Spanish Ministerio de
Educaci{\'o}n y Ciencia (MEC) and FEDER funds under grant ESP
2003-04957, and from the Secretar\'{\i}a de Estado de Pol\'{\i}tica
Cient\'{\i}fica y Tecnol\'ogica/MEC under grant AYA 2003-07584.
\end{acknowledgement}

\end{document}